\documentclass[prb,a4paper,twocolumn,showpacs,amsfonts,amssymb,amsmath]{revtex4-1}%
\usepackage{graphics}
\usepackage{dcolumn}
\usepackage{bm}
\usepackage{subfigure}
\usepackage{epsfig}

\usepackage{amsmath}
\usepackage{amsfonts}
\usepackage{amssymb}
\usepackage{arcs}

\begin{document}

\title{Path to collapse for an isolated N\'eel skyrmion}

\author{S. Rohart}   \email{stanislas.rohart@u-psud.fr}
                     \affiliation{Laboratoire de Physique des Solides, CNRS, Univ. Paris-Sud, Universit\'e Paris-Saclay, 91405 Orsay Cedex, France}
\author{J. Miltat}   \affiliation{Laboratoire de Physique des Solides, CNRS, Univ. Paris-Sud, Universit\'e Paris-Saclay, 91405 Orsay Cedex, France}
\author{A. Thiaville}\affiliation{Laboratoire de Physique des Solides, CNRS, Univ. Paris-Sud, Universit\'e Paris-Saclay, 91405 Orsay Cedex, France}

\date{\today}

\begin{abstract}
A path method is implemented in order to precisely and generally describe the collapse of isolated skyrmions in a Co/Pt(111) monolayer, on the basis of atomic scale simulations. Two collapse mechanisms with different energy barriers are found. The most obvious path, featuring a homogeneous shrinking gives the largest energy, whereas the lowest energy barrier is shown to comply with the outcome of Langevin dynamics under a destabilizing field of 0.25 T, with a lifetime of 20~ns at around 80~K. For this lowest energy barrier path, skyrmion destabilization occurs much before any topology change, suggesting that topology plays a minor role in the skyrmion stability. On the contrary, an important role appears devoted to the Dzyaloshinskii-Moriya interaction, establishing a route towards improved skyrmion stability.
\end{abstract}


\maketitle

\section{Introduction}

The Dzyaloshinskii Moriya interaction (DMI)~\cite{dzyalo1965,moriya1960} promotes a non-collinear order, and, thus, is at the origin of new states of magnetism, neither ferro- nor antiferromagnetic. They differ from structures arising from competing exchange interactions~\cite{yoshimori1959,villain1959} by a chirality that is fixed. The stabilization of skyrmions~\cite{bogdanov2001}, which correspond to whirling configurations that are topologically different from the uniform ground state, is one of its most fascinating consequences. First found in arrays which are the ground state of the system~\cite{yu2010,heinze2011}, they can also be obtained as solitons, i.e. localized excited states above the energy of the ferromagnetic ground state~\cite{kiselev2011prl,rohart2013}. In that case, promising applications in spintronics are foreseen such as ultra dense memories~\cite{fert2013} or logic devices~\cite{zhang2015}. Recently, significant progress has been made in the control of isolated skyrmions including nucleation and motion under spin polarized currents~\cite{romming2013,jiang2015,woo2016}.

The stability of skyrmions is a major issue regarding their possible use in room-temperature devices.
As they are distinct from the uniform state by a different topology, an exceptional stability may be expected~\cite{fert2013}.
Topology, described by the topological number $S$, is intimately linked to a continuous description of magnetism, in the framework of micromagnetics~\cite{feldkeller1965}.
As integer values of $S$ only are possible, no continuous path linking different topological states  exists, hence the topological stability.
In the discrete atomic description, however, topological considerations do not hold and in particular no univoque equivalent of $S$ exists.
On the other hand, topologically forbidden transitions in magnetism are experimentally observed, e.g. in the case of magnetic bubble collapse~\cite{Malozemoff79} or vortex core reversal~\cite{shinjo2000,thiaville2003}, as well as for skyrmions~\cite{romming2013,jiang2015}.
Thus, the details of topological transitions have to be studied in order to understand how the topological paradox is lifted, and to evaluate the intrinsic stability of skyrmions.
In the continuous micromagnetic description within a three-dimensional world, the topological transition is realized by the injection of a topological singularity called Bloch point, where the continuity of magnetism is broken~\cite{feldkeller1965}.
For skyrmions, nucleation and annihilation have been observed in various simulations~\cite{sampaio2013,verga2014,zhou2015,Hagemeister2015,Bessarab2015,rozsa2016} under thermal fluctuations or spin transfer torque excitations, but a detailed mechanism is still missing to evidence the role of topology in the stability.
Whereas a Bloch point cannot, strictly speaking, exist in two-dimensional samples, an equivalent process has been mentioned~\cite{sampaio2013,zhou2015} but it is not clear how it relates to the intermediate configuration at the energy barrier.

In this paper, we investigate by atomic scale calculations the path linking an isolated skyrmion to the ferromagnetic state and its consequences on the skyrmion stability.
Using static calculations to determine the collapse path, we describe the collapse process and show the importance of DMI in the stability. Results are further confirmed by the study of thermally induced collapse, using Langevin dynamics.
Rather than using arbitrary parameters, we focus our calculations on skyrmions in a Co monolayer on Pt(111).
This is one of the most promising systems that are studied in order to control isolated skyrmions.
Indeed, while it presents a large DMI~\cite{freimuth2014,yang2015,belmeguenai2015} the strong out-of-plane magnetic anisotropy and Heisenberg exchange prevent intrinsic destabilization of the ferromagnetic order, which is key to enabling isolated skyrmions.
In addition, we develop a very fast shortcut to minimum energy path calculations in the case where a reaction coordinate along the path is known.

\section{Magnetic medium and skyrmion description}

The magnetic layer is described at the atomic scale, by a set of classical spins $\mathbf{S}_i$ on an hexagonal lattice realizing an epitaxial Co monolayer on a Pt(111) substrate, with a site-to-site distance $a=2.51$~\AA. The energy is given by
\begin{equation}\begin{split}\label{eq_energyDM}
    E = \sum_{<i,j>}\left[-J\mathbf{\hat{s}}_i.\mathbf{\hat{s}}_j+\mathbf{d}_{ij}.\left(\mathbf{\hat{s}}_i\times\mathbf{\hat{s}}_j\right)\right]-
    \sum_ik\left(\mathbf{\hat{s}}_i.\mathbf{\hat{z}}\right)^2\\-\frac{\mu_0}{8\pi}\sum_{i,j\neq i}\frac{3\left(\mathbf{S}_i.\mathbf{u}_{ij}\right)\left(\mathbf{S}_j.\mathbf{u}_{ij}\right)-\mathbf{S}_i.\mathbf{S}_j}{r_{ij}^3}
    -\mu_0\sum_i\mathbf{S}_i.\mathbf{H}.
\end{split}\end{equation}
where $\mathbf{\hat{s}}_i=\mathbf{S}_i/\|\mathbf{S}_i\|$. The two first terms are respectively the Heisenberg and Dzyaloshinskii-Moriya interaction with respective constant $J$ and $\mathbf{d}$. Summation is performed on the first neighboring pairs $<i,j>$, which accurately describes Co/Pt(111)~\cite{dupe2014}. For a thin film, $\mathbf{d}_{ij}=d(\mathbf{\hat{u}}_{ij}\times \mathbf{\hat{z}})$ \cite{moriya1960,fert1990}, with $\mathbf{\hat{u}}_{ij}$ the unit vector between sites $i$ and $j$ and $\mathbf{z}$ the normal to the plane. The third term is the uniaxial anisotropy, with constant $k$, the fourth term is the dipolar coupling and the last term is the Zeeman energy in field $\mathbf{H}$. The parameters are: $\mu_{at}=\|\mathbf{S}_i\|=2.1$~$\mu_B$/atom~\cite{sipr2007,gambardella2003,nakajima1998,moulas2008}, $J=29$~meV/bond~\cite{sipr2007}, $d=1.5$~meV/bond~\cite{freimuth2014,yang2015,belmeguenai2015}, $k=0.4$~meV/atom~\cite{gambardella2003,moulas2008} (including the shape anisotropy~\cite{rohart2007}, the effective anisotropy is 0.276~meV/atom in good agreement with literature~\cite{gambardella2003,moulas2008}). The sample is limited by free boundary conditions and the size has been chosen so that an isolated skyrmion is not affected by the edges (no morphology nor energy changes seen for larger calculation box sizes).


The ground state of our system is ferromagnetic which means that DMI is not sufficient to destabilize the collinear magnetic order, in agreement with experimental observations (the onset of ferromagnetic order instability is  $d_c=2.08$ meV here). The ferromagnetic order remains stable for all temperatures up to the Curie temperature, estimated to be 375~K (using Monte Carlo simulations - not shown). However, when a skyrmion is introduced it remains as a metastable state. The skyrmion (see Fig.~\ref{fig_chemin}a) has a radial symmetry with a hedgehog configuration (in-plane magnetization pointing along the radial direction). Conceivably, skyrmions could be centered on a lattice site or have their core split-up over a lattice triangle. The latter structure is found to have the smallest energy, although the energy difference proves minute. For the present parameters and in zero field, its energy is 485~meV, with a diameter of 4.6 nm in quite good agreement with analytical predictions~\cite{rohart2013}.

\section{Skyrmion collapse path}

Deciphering the collapse mechanism implies a search for the easiest path that links the skyrmion to ferromagnetic states. This has been achieved, although not at the atomic scale, in the case of magnetic vortex core reversal~\cite{thiaville2003} - a problem quite close to ours - and good agreement with the thermal fluctuation method reached ~\cite{dittrich2004}. To determine the minimum energy path between two stable configurations on the multidimensional energy surface~\cite{Henkelman2000} 21 images, from the skyrmion to the uniform state, are energy-minimized simultaneously. Relaxation occurs in the plane orthogonal to the tangent between successive images~\cite{annexe}. An elastic force along the tangent is also added to ensure equidistance between images, relying on a geodesic metric~\cite{Bessarab2015,annexe}.  Given the complexity of the energy surface and the large number of degrees of freedom, several stable paths may be obtained depending on the initial guess.
\begin{figure}[ht]
\includegraphics[width=\columnwidth]{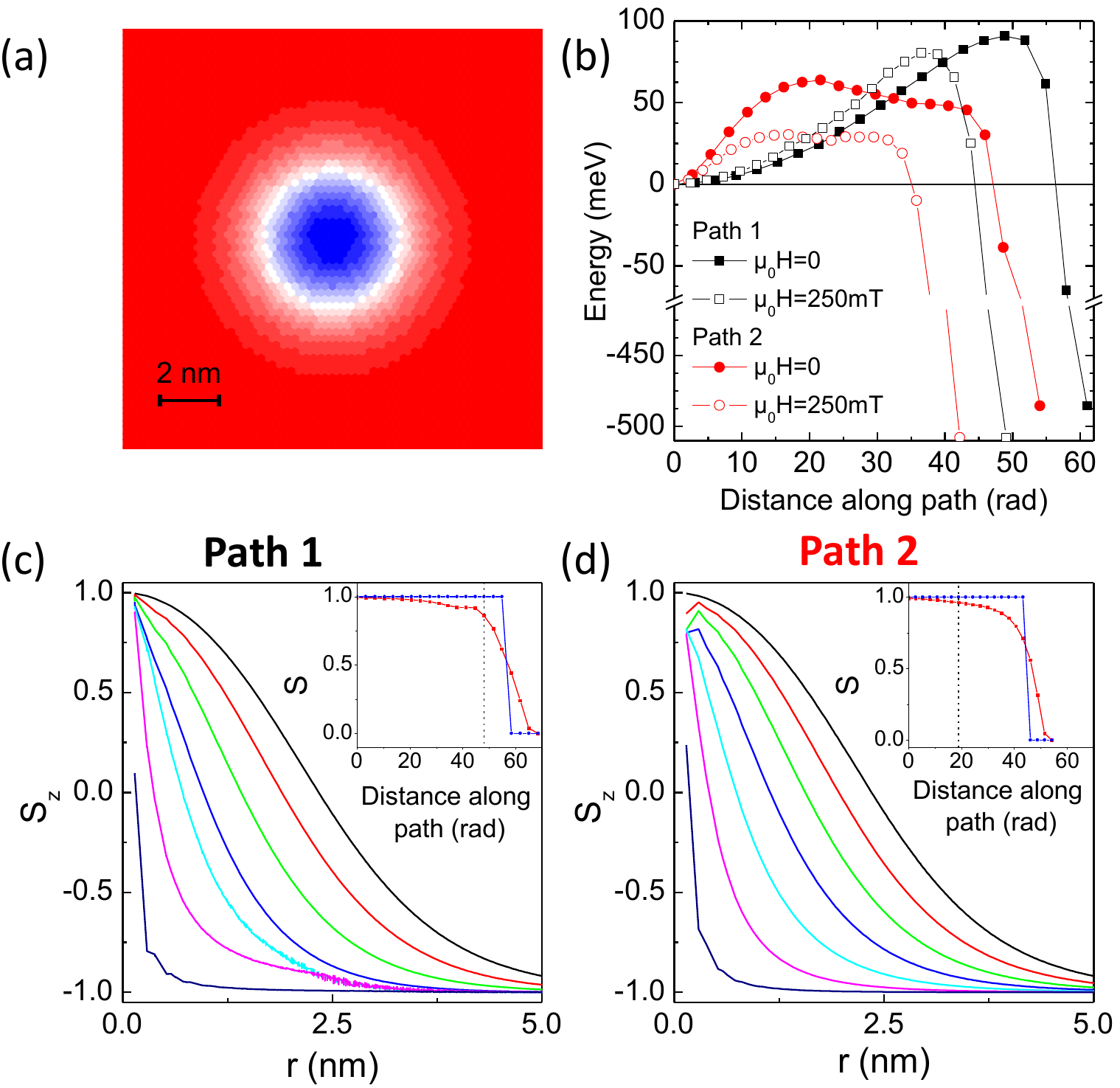}
\caption{(color online) (a) Skyrmion configuration in zero field. The color code represents the perpendicular spin component from blue (upward) to red (downward) through white (in-plane). (b) Energy variation along the collapse path. The distance along the path corresponds to the geodesic metric~\cite{Bessarab2015,annexe}. (c-d) Variation of the perpendicular spin component as a function of the radial distance for successive configurations along the collapse path (starting from the skyrmion state, the profiles are plotted for every third configuration along the path) in zero magnetic field. In (c) the path only involves diameter reduction (path~ 1). In (d) the path involves both skyrmion diameter reduction and skyrmion center spin rotation (path~2). In the insets, the variation of the skyrmion number with two definitions (see text) are plotted (the dotted line indicates the position of the energy maximum). }
\label{fig_chemin}
\end{figure}

Two paths have been identified (see Fig. \ref{fig_chemin}). Path 1 is characterized by a progressive skyrmion diameter reduction down to zero (Fig. \ref{fig_chemin}b,c), the texture being almost self-similar along the path. Such a path may appear as natural and is the only one considered in previous studies~\cite{Bessarab2015,verga2014}. Path 2 appears more sophisticated, involving, on top of an overall skyrmion diameter reduction as in path 1,  a large rotation of the center spins in the radial direction and in a sense opposing the DMI favored chirality, that can be viewed as a coherent excitation of the most central spins (Fig. \ref{fig_chemin}b,d). 
For path 1 the energy increase is progressive and the maximum (saddle point) is reached when the skyrmion is compressed down to the few last center lattice sites (Fig. \ref{fig_chemin}b). For path 2, the initial energy rise proves faster but the saddle point is obtained much earlier, showing that the rotation of the center spins efficiently contributes to the skyrmion destabilization (Fig. \ref{fig_chemin}b).
Path 2 proves to be the lowest energy one with an energy barrier of 64 meV compared to 90 meV for path 1. Under a finite (destabilizing i.e. oriented opposite to the skyrmion core) field of 250 mT, path 2 is also found to be the easiest one with an energy barrier falling to 30 meV.
Path 2 also exhibits a larger susceptibility to field as compared to path 1 (Fig.~\ref{fig_chemin}b).

\begin{figure}[ht]
\includegraphics[width=\columnwidth]{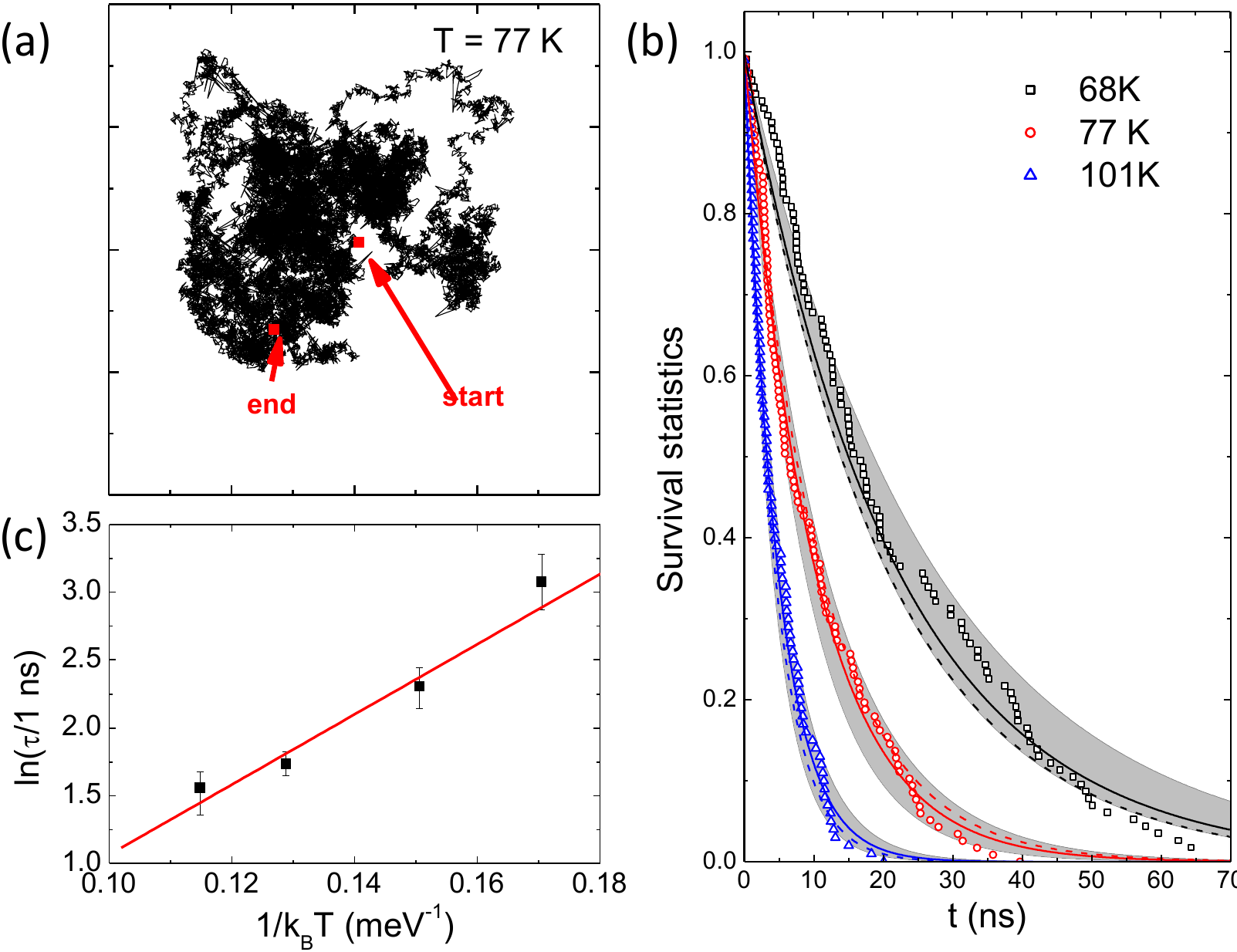}
\caption{(color online) (a) Trajectory of a skyrmion submitted to thermal noise at 77 K, from start (initial position) to end (skyrmion collapse). The frame represents the $200\times200$ nm$^2$ calculation box. Skyrmion collapse occurs here after 28 ns. (b) Survival statistics versus time for three different temperatures. The full line is the $\exp(-t/\tau)$ model (with $\tau$ the mean collapse time, function of temperature) and the grey shaded area represents the uncertainty on $\tau$.  (c) Arrhenius plot of the mean lifetime $\tau$ versus temperature. The line is a fit to the data leading to activation energy and attempt time.}
\label{fig_thermique}
\end{figure}

These statics results are confronted to thermally induced collapse using Langevin dynamics. Although this latter method hardly allows for a description of the collapse mechanism, it is free from any hypothesis on the collapse path. It involves a large amount of calculations to account for the process statistics, so that the time scale that can reasonably be explored is only a few tens of nanoseconds. In order to observe a significant number of skyrmion collapse events, stability is only considered under a 250~mT out-of-plane destabilizing field and calculations have been conducted around 80~K, with no less than 100 collapse events per temperature.

As a first outcome of these calculations, skyrmion diffusion is observed (Fig.~\ref{fig_thermique}a): the skyrmion being a soliton and its energy independent of position, free random walk is indeed expected~\cite{einstein1905,schutte2014}. After some time, the skyrmion vanishes suddenly.
The survival statistics obtained from the collapse times distribution (Fig.~\ref{fig_thermique}b) is well fitted by a simple exponential decay for all four considered temperatures, suggesting that only one markovian process is involved. Of course, the higher the temperature, the shorter the mean life time $\tau$. An activation energy is extracted\cite{Hagemeister2015,rozsa2016} from the Arrhenius law $\tau = \tau_0\exp(\Delta E/k_BT)$. In our case, although the agreement is not perfect, indicating that such a law may be oversimplified (for example, no temperature dependence of $\tau_0$ is included), it sets the order of magnitude of the energy barrier to $\Delta E = 26\pm 4$~meV and $\tau_0 = 0.22\pm0.1$~ns (Fig.~\ref{fig_thermique}c), in good agreement with the lowest energy barrier found using the path analysis.


\section{Discussion}
\subsection{Role of topology}

In order to assess the role of topology in the skyrmion stability, a topological number $S$ along the path needs to be defined. In the original work of Feldtkeller that deals with continuous textures~\cite{feldkeller1965}, two equivalent definitions are given. The \emph{geometrical} definition measures the proportion of the unit sphere (the order parameter space) covered by the spin texture projected onto it. This quantity is \emph{mathematically} expressed as $S=(4\pi)^{-1}\int\mathbf{m}.(\partial \mathbf{m}/\partial x\times\partial \mathbf{m}/\partial y)dxdy$. Both definitions can be extended to a discrete set of spins, either (geometrical $S$) by tiling the unit sphere with oriented spherical triangles whose apices coincide with  the projections of neighboring spins on the same unit sphere~\cite{thiaville2003}, or (finite differences $S$) by substituting discrete to ordinary partial derivatives in the expression of $S$~\cite{annexe}. These two definitions, however, prove not anymore equivalent: whereas the geometrical description still yields integer $S$ values, the finite differences formulation leads to various values and thus may describe a continuous transition from the skyrmion to the ferromagnetic state. In Fig.~\ref{fig_chemin}c,d, the variation of $S$ along the path is shown for both definitions.

\begin{figure}[ht]
\includegraphics[width=\columnwidth]{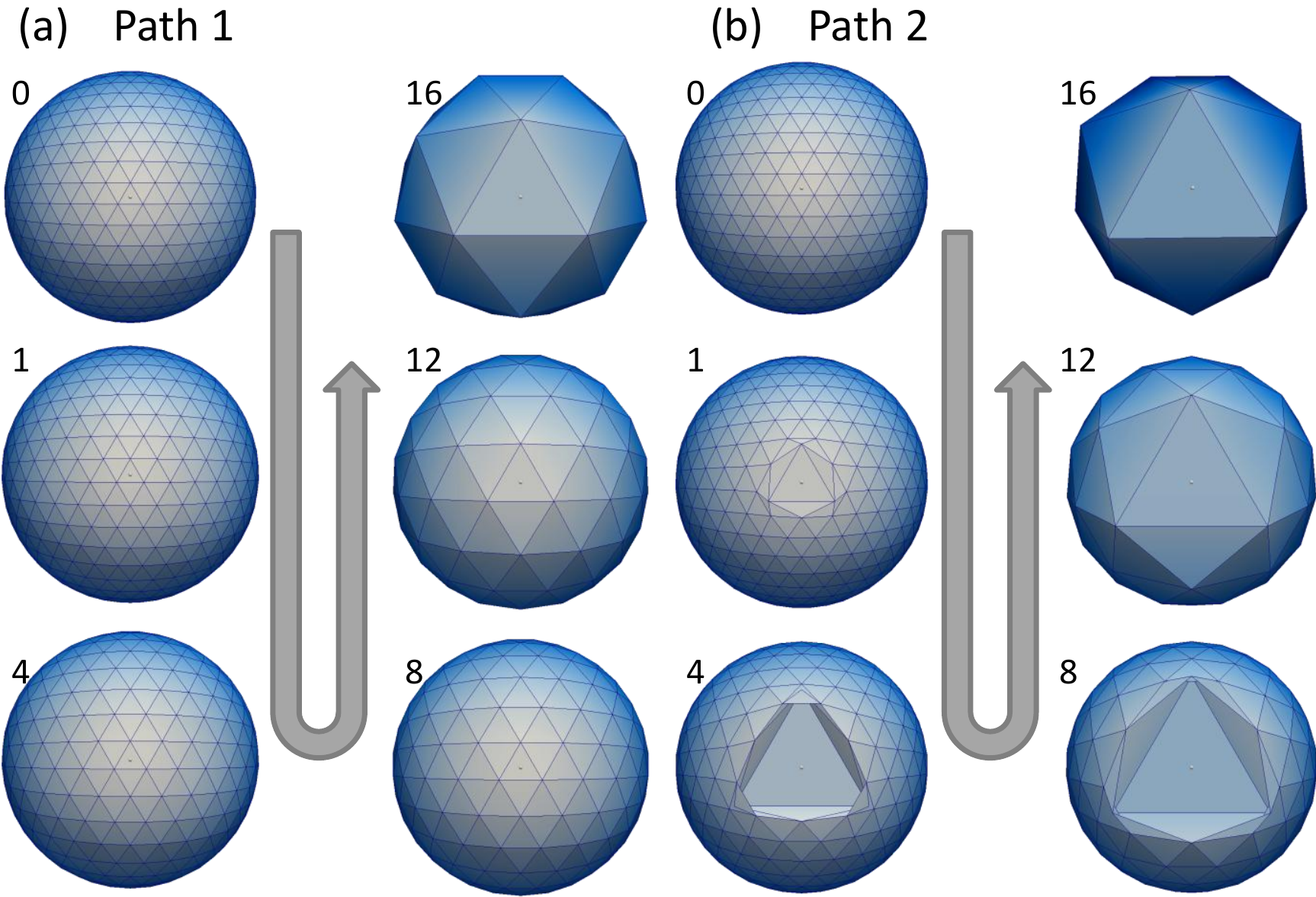}
\caption{(a) Mapping on the unit sphere, viewed from well above the top hemisphere (skyrmion core orientation), of six magnetic textures belonging to path 1 for $H_a = 0$.
To each node on the sphere corresponds a spin with origin at the sphere center and end point at the node.
The drawn net is made out of straight lines (in 3D space) connecting points corresponding to nearest neighbors.
Numbers refer to the various images along the path, starting with 0, corresponding to the skyrmion at equilibrium, ending with 20, corresponding to the uniform state. Image 16 corresponds to the energy maximum.
Path 1 is characterized by an homogeneous opening-up of the spin distribution that matches a gradual shrinking of the skyrmion radius in the physical space.
The color scale represents reflectivity towards the observer, for a light source that is vertically situated with a small tilt towards
page top.
(b) Same as (a) for path 2. Here, the energy maximum occurs at image 8. Together with a gradual opening of the spin distribution, path 2 is primarily characterized by a strong fanning-out of the three core spins.
Note that, from image 2 onwards, some nearest-neighbor links are hidden, and triangles significantly inclined with respect to the
local sphere tangential plane.}\label{fig_sphere}
\end{figure}

Mapping the spins on the unit sphere (order parameter space) provides a simple and efficient representation of the texture topology along a path. In Fig. \ref{fig_sphere}, only the top hemisphere is visible (along the skyrmion core orientation). In the skyrmion state, the regular tiling of the sphere underlines a smooth texture for the few-nanometer-diameter skyrmion. Along path 1 (Fig. \ref{fig_sphere}a), the gradual swelling of the triangular spin lattice pattern simply represents the regular skyrmion shrinking process. In image 16, which corresponds to the energy maximum, only a few spins still have a component along the top pole direction. Markedly different is the onset of path 2 (Fig. \ref{fig_sphere}b). From the very beginning, a very local, and symmetrical,  distortion of the three most inner spins is allowed for. The distortion grows with image number, retaining the three-fold symmetry. In image 8, which corresponds to the energy maximum, the number of spins with a component along the to hemisphere direction is already strongly reduced, although still larger than the number of spins with the same property in image 16 of path 1. Adding a moderate skyrmion destabilizing uniform field only marginally modifies these patterns. The topology change occurs when the top hemisphere is fully depleted, and when three spins (those at the center) lie on the equator, thus are in-plane, pointing radially. This last configuration is similar, in 2D, to the horizontal cut through the Bloch point observed in thicker samples~\cite{thiaville2003}. The maximum in energy is thus obtained before (or even well before -path 2-) the topological transition\cite{annexe}.  A trivial analogy is to consider the magnetization textures as elastic nets through which one tries to extract a balloon. Path 1 strains the whole net whereas Path 2 concentrates the strain on a single mesh cell.

\subsection{Energy considerations}

\begin{figure}[ht]
\includegraphics[width=\columnwidth]{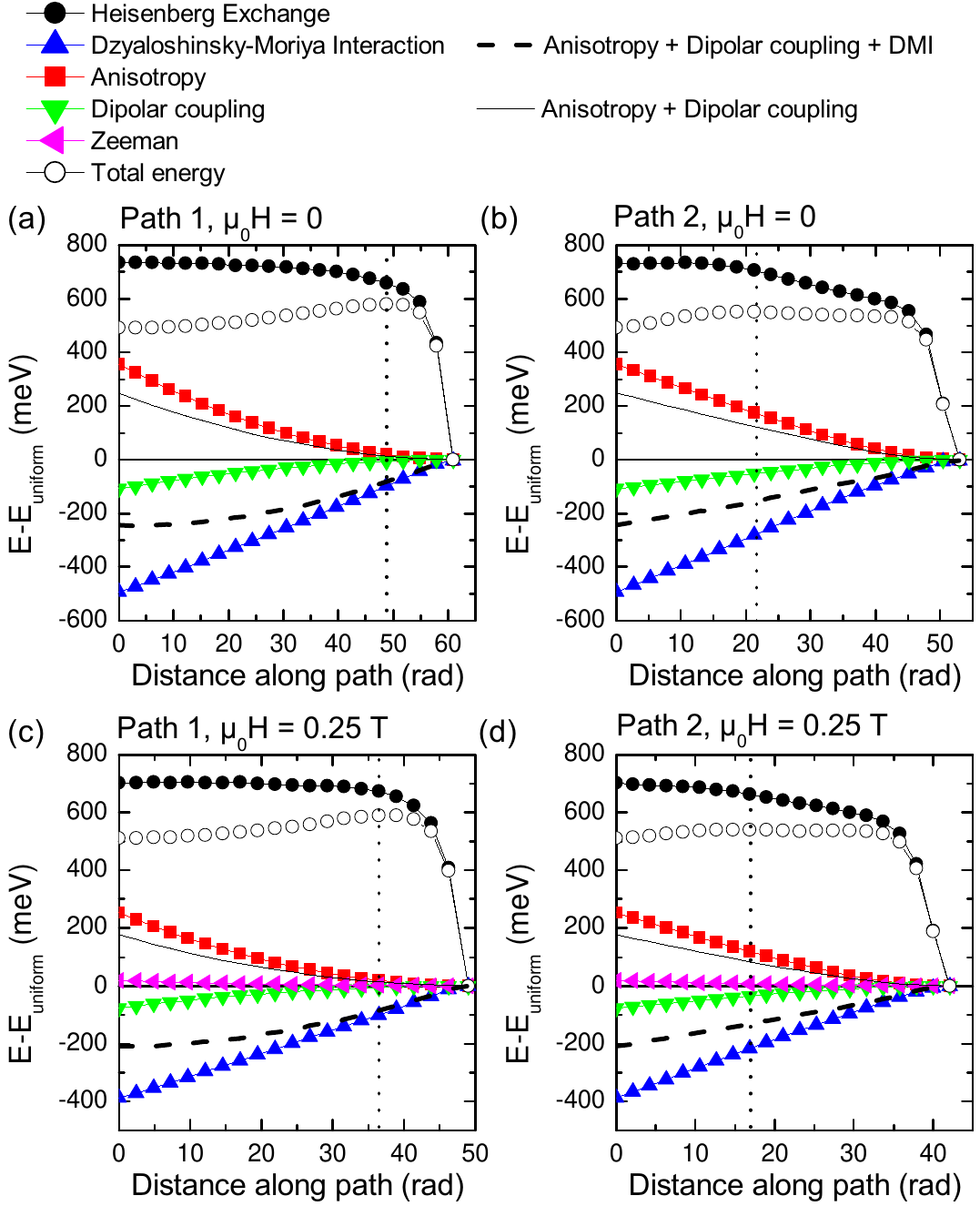}
\caption{Variation of the various energy contributions along path 1 and path 2 and at 0 and 0.25~T. The vertical dotted lines indicate the position of the total energy maximum (energy barrier).}\label{fig_energy}
\end{figure}

We now investigate the contribution of the different energy terms to the barrier, as shown for every energy components along paths 1 and 2 in Fig. \ref{fig_energy}, both for $H_a = 0$ and $H_a = -0.25$ T. The variations of the dipolar coupling and anisotropy, although opposite in sign, are equivalent, which indicates that dipolar coupling mainly acts as a shape anisotropy (0.124~meV/atom) and that the sum of both of them can be considered as an effective anisotropy.
A common feature to all graphs is the large relative variation along the path of the (effective) anisotropy and Dzyaloshinskii-Moriya energies as compared to the exchange or Zeeman energies.
It even appears that, in the approach to the maximum energy along a given path, the slope of the total energy is governed by the behavior of the sum of the effective anisotropy and Dzyaloshinskii-Moriya energies.
Note in this respect that this sum monotonously increases towards the barrier top for path 2, whereas it first goes
through a minimum for path 1.
In contradistinction, the exchange energy is almost constant in this part of the path (see appendix).


\section{Increasing skyrmion lifetime}

From these results, it is possible to estimate the lifetime of an isolated skyrmion in zero magnetic field, using $\tau_0=0.22$~ns and $\Delta E = 64$~meV (note that the error bar on $\tau_0$ is not taken into account as it doesn't change the order of magnitude of the lifetime). It implies that whereas almost infinite at 4~K, the lifetime is only 4~$\mu$s at 77~K and 3~ns at room temperature. Whereas zero temperature calculations do show for material parameters such as above the possibility to stabilize isolated skyrmions, the latter can only be studied at really low temperature to avoid thermal  collapse.

\begin{figure}[ht]
\includegraphics[width=\columnwidth]{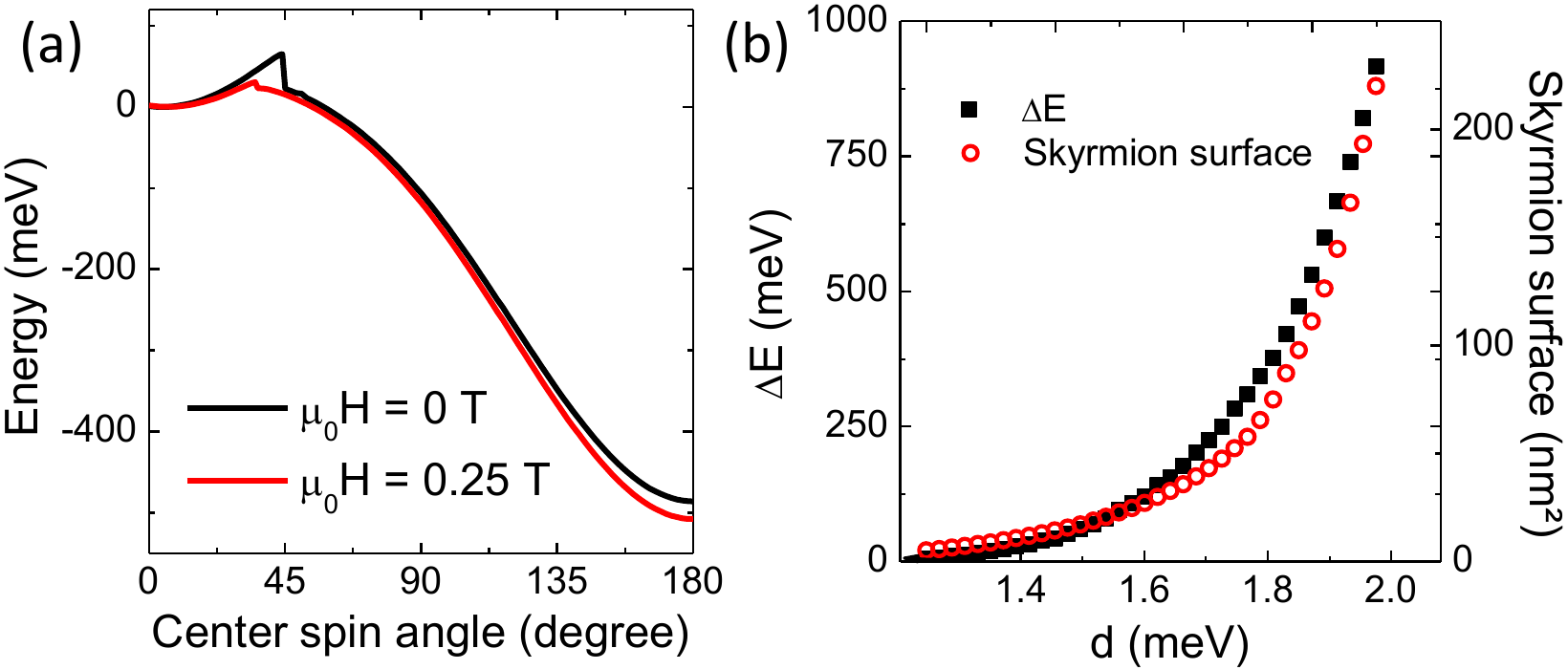}
\caption{(color online) (a) Skyrmion energy variation as a function of an imposed spin angle for the center spins for zero and 0.25 T field. The energy reference is the skyrmion energy. (b) Collapse activation energy and skyrmion surface as a function of the DMI strength calculated using the same method as in (a).}
\label{fig_adhoc}
\end{figure}

\begin{figure}[ht]
\includegraphics[width=\columnwidth]{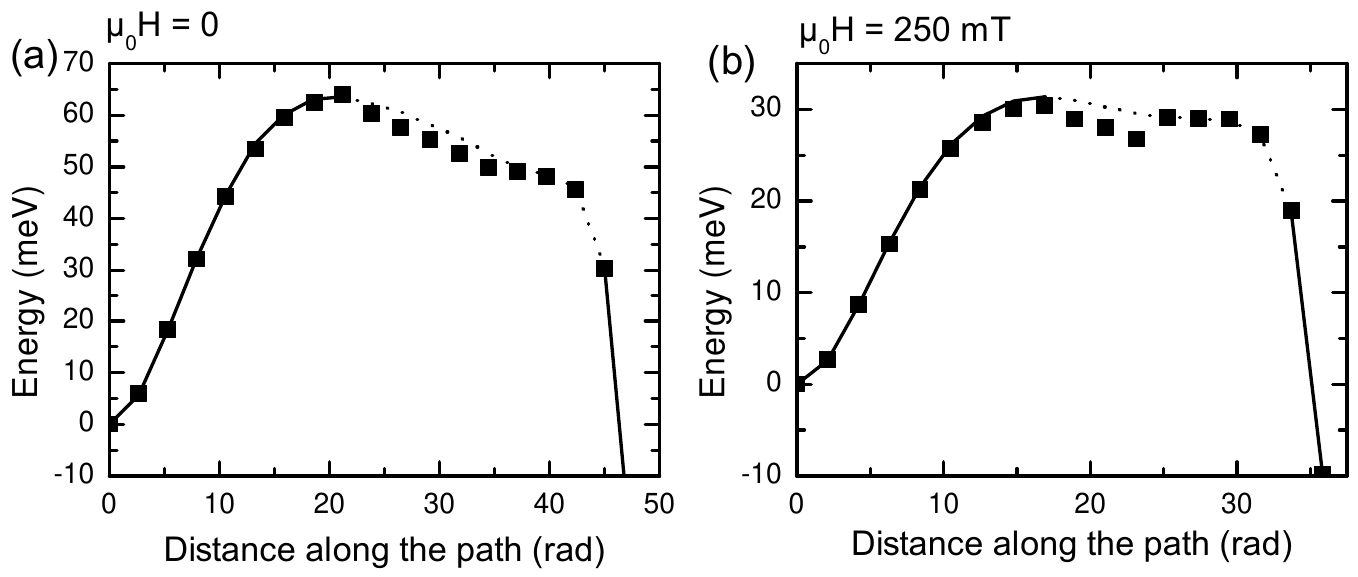}
\caption{Energy variation along the collapse path for \textit{ad-hoc} method (lines) and path method for path 2 (square dots) in (a) zero and (b) 250~mT field and for $d=1.5$~meV. The full line indicates energies for configurations that are stable for the imposed center spin angle. The dotted line corresponds to the energy drop that occurs in the calculation (see Fig. 3a) above a certain imposed center spin angle, and thus to configurations that are not stable with respect to the imposed angle.}
\label{fig_comparaison_adhoc/path}
\end{figure}

The poor stability observed here is due to the monolayer thickness on the one hand but also, on the other hand, to too low a DMI. Indeed, whereas a moderate DMI is required to avoid a spontaneous skyrmion lattice, the DMI value used here is only 72\% of the maximum possible value allowing for isolated skyrmions. To study the influence of the DMI on skyrmion stability, we have arbitrarily varied its magnitude, $d$. Previous methods could not be used for calculation time reasons so that we need to introduce an \textit{ad-hoc} method. Having identified the crucial role of the spins at the skyrmion center, we manipulate them by imposing their orientation thus mimicking the lowest energy path, and then obtain the spin configuration at equilibrium for the other sites via energy minimization. The validity of this approach compared to the path method is checked by verifying that, for the center spins, due to the symmetry, the effective field is aligned with the tangent between successives images. By varying the center angle from 0$^\circ$ (skyrmion state) to 180$^\circ$ (uniform state), the results prove quite close to the easiest path found previously (Fig.~\ref{fig_adhoc}a), but with a much reduced calculation time. We observe that the reaction to a small change of the center spin tilt is a skyrmion diameter reduction and an energy increase. Above a critical tilt angle, the configuration undergoes a brutal change synonymous of skyrmion collapse (the non-collinear situation remaining only in the vicinity of the center spins) and an energy drop. Further increase of the tilt angle describes a progressive transition to the ferromagnetic state. Both before and after the critical angle, the energy curve versus metrics is almost undistinguishable from the more rigorous path method calculation (the maximum energy is overestimated by about 3~\%). The brutal configuration change corresponds to a large jump along the path (see Fig.~\ref{fig_comparaison_adhoc/path}).
The variation of the collapse activation energy as a function of $d$, shown in Fig. \ref{fig_adhoc}b, displays a strong non-linear dependence. It appears that a modest increase of DMI significantly improves the stability. As an example, an increase of 0.2 meV ($\approx$ 13~\%) results in $\Delta E = 220$~meV (240~\% increase). In that case the skyrmion lifetime becomes 13~h at 77~K and 1~$\mu$s at room temperature. Such a dependence can be interpreted by an increase of the skyrmion diameter (for $d = 1.7$~meV, the diameter increases to 7.3 nm), which increases the amount of spins to be reversed. However, the correlation between $\Delta E$ and the skyrmion surface is not exact as the energy barrier results from a complex balance of energies, and as also early guessed because the isolated skyrmion size diverges at the critical DMI value.

\section{Conclusion}

In conclusion, we have studied the collapse of isolated skyrmions and identified a non-trivial easiest mechanism as a progressive diameter reduction combined with a rotation of spins at the skyrmion center. For such a mechanism, the destabilization occurs much before any topology change, suggesting that topology plays a minor role, with respect to the micromagnetic energy balance. On the contrary, the DMI energy is shown to play an important role. Despite the rather poor skyrmion stability found in the monolayer Co/Pt(111), our study opens a route to improve it. A modest DMI increase, using interface engineering combining under and top layers with opposite sign DMI, as a Pt/Co/Ir~\cite{hrabec2014,yang2015} or Pt/Co/MgO~\cite{boulle2016}, could be a solution toward room temperature stability for nanometer sized skyrmions.

\section*{Appendix}

The near-constancy of the exchange energy can be understood simply.
On the one hand, in the continuous formulation it is immediate to see that exchange energy is invariant under
spatial expansion [$\vec{m}(x,y) \rightarrow \vec{m}(\lambda x, \lambda y)$].
On the other hand, in the discrete formulation, the energy cost of non-parallel nearest neighbor spins is
\begin{equation}
E_{\mathrm{exc},ij} = J \left( 1-\mathbf{\hat{s}}_i \cdot \mathbf{\hat{s}}_j \right) = J \left( 1-\cos \theta_{ij} \right).
\end{equation}
The last formula has a geometrical interpretation: the area on the unit sphere of the cone $C_{ij}$ of axis
$\mathbf{\hat{s}}_i$ and with $\mathbf{\hat{s}}_j$ on its generatrix is
\begin{equation}
C_{ij}=2 \pi  \left( 1-\cos \theta_{ij} \right).
\end{equation}
If the distribution of the spins on the unit sphere is regular, one then expects that
\begin{equation}
C_{ij} \approx 3 \left( T^+_{ij} + T^-_{ij} \right),
\end{equation}
where $T^{\pm}_{ij}$ are the two spherical triangles that have $\mathbf{\hat{s}}_i$ and $\mathbf{\hat{s}}_j$ as apices.
Instead of the approximate equality, we can introduce a numerical factor $f$, whose value is
$2 \pi / (3 \sqrt{3})$ for infinitely small triangles.
By summation over the bonds one gets
\begin{equation}
E_\mathrm{exc} = \sum_{<i,j>} E_{\mathrm{exc},ij} = \frac{9 f J}{2 \pi} \sum T.
\end{equation}
If the sphere is covered once, then $\sum T = 4 \pi$ so that a constant exchange energy is found, with a value
\begin{equation}
E_\mathrm{exc}= 18 f J = 4 \pi \sqrt{3} J,
\end{equation}
the latter equality stemming from using the value of $f$ for regular small triangles.
Numerically, with $J=29$~meV one finds $E_\mathrm{exc}= 631$~meV, of the same order as the numerical results.
The (cumbersome) expression of $f$ for finite size equilateral spherical triangles shows that it decreases
as the triangle size increases, hence the observed decrease of the exchange energy along the path.
This result is in contrast with that of the continuous approach, where it was proved \cite{Belavin1975} that
$E_\mathrm{exc} \ge 8 \pi A_\mathrm{2D}$, with $A_\mathrm{2D}= J \sqrt{3} / 2$ is the micromagnetic
exchange constant for the corresponding 2D medium.

\begin{acknowledgments}

This work was supported by Agence Nationale de la Recherche under contract ANR-14-CE26-0012 ULTRASKY.
We thank members of this consortium, especially A. Fert and J. Sampaio, for stimulating discussions.
\end{acknowledgments}

\newpage

\section*{Supplementary information}

\subsection{S.1 Minimum energy path method}

The aim is to find a minimum energy path (MEP) between two stable (or metastable) energy minima on a multidimensional energy surface. We assume that such a path may be approximately built through a continuous transformation of the spin configuration between the configurations pertaining to energy minima A and B and sampled as a set of spin configurations referred to as "images", labeled $n=1,N$.
Each image is a set of $\mathcal{N}$ spins, $i=1,\mathcal{N}$.
For the results shown, we used $N= 21$, and $\mathcal{N}= 36246$.
Let first the tangent to the path be defined as the set of unit vectors:
\begin{equation}\label{A1}
    \hat{t}_i^{n+1/2}=\frac{\mathbf{S}_i^{n+1}-\mathbf{S}_i^n}{\|\mathbf{S}_i^{n+1}-\mathbf{S}_i^n\|}
\end{equation}

For the path to represent a chain of metastable minima, energy minimization needs to be performed within the hyperplane normal to the tangent to the path. This precept applied to a set of classical Heisenberg spins implies that energy minimization may solely result from spin rotation in a plane locally defined by $\mathbf{\hat{s}}_i$ and $\hat{\eta}_i = \mathbf{\hat{s}}_i\times\mathbf{\hat{t}}_i$ (we omit the $n$ index for legibility). This condition may be viewed as a constraint to energy minimization. The spin velocity in the $(\mathbf{\hat{s}},\hat{\eta})$  plane thus reads:
\begin{equation}\label{A2}
    \Delta \mathbf{S}_i \propto (\mathbf{h}_i.\hat{\eta}_i)\hat{\eta}_i
\end{equation}
where, $\mathbf{h}_i$  is the effective field acting on spin $\mathbf{S}_i$. Equation \ref{A1} implies that the tangent vector is undefined for spins that remain constant within two successive images. In the present implementation of the MEP method, instead of leaving such spins untouched, it is preferred to allow for some unconstrained rotation defined as the usual damping-dominated spin relaxation according to the Landau-Lifshitz-Gilbert equation of spin motion, namely:
\begin{equation}\label{A3}
    \Delta \mathbf{S}_i \propto \mathbf{\hat{s}}_i\times(\mathbf{h}_i\times\mathbf{\hat{s}}_i)
\end{equation}
Eqs. \ref{A2} [or \ref{A3} whenever the tangent vector is undefined]  applied to all images except the end ones allows the successive images to reach their local metastable configuration.
	Upon relaxation, however, images may move with respect to each other along the path, pile-up against end-images, or, more simply, fail to remain roughly equi-spaced along the path, all facts detrimental to a precise determination of the barrier height at saddle point. In the nudged elastic band (NEB) method an elastic force between images is added in order to maintain equidistance between successive images \cite{Henkelman2000,Dittrich2002}. To this end, a metric needs to be defined. In the present implementation, the geodesic \cite{Bessarab2015} metric relies on the arc lengths on the unit sphere spanned by the extremities of corresponding spins belonging to two successive images, namely:
\begin{equation}
\mathcal{L}_{n-1,n} = \sqrt{\sum_{i=1}^{\mathcal{N}}\left(\mathbf{\hat{s}}_i^n\mathbf{\hat{s}}_i^{n-1}\right)^2},
\end{equation}
therefore, the metric is in radians.

In the NEB method \cite{Henkelman2000}, relaxation in the hyperplane normal to the path and elastic translation of the images along the path are performed concomitantly. Other optimizing schemes have been proposed, e.g. Ref. \onlinecite{Samanta2013}, implying some mixing and reparametrization performed in succession. In the present implementation, a sole constrained relaxation is often found preferable, as long as the input path results from an educated guess, with equispaced images. The end results also prove little sensitive to whether or not the tangents to path are strictly tangent to the unit sphere, as advocated in \cite{Bessarab2015}.

Lastly, Eq.\ref{A2} implies that no magnetization motion takes place within an image whenever the effective field $\mathbf{h}_i$ belongs to the  $(\mathbf{\hat{s_i}},\mathbf{\hat{t_i}})$ plane. Such conditions may arise along high symmetry planes within a magnetic texture, leading to a local "quenching" of the path.

\subsection{S.2 Topological number for discrete description of magnetism}

In the continuous formulation of magnetism, the topological number is described by

\begin{equation}
S=\frac{1}{4\pi}\int\mathbf{m}.\left(\frac{\partial \mathbf{m}}{\partial x}\times\frac{\partial \mathbf{m}}{\partial y}\right)dxdy
\end{equation}

To extend this definition to a discrete lattice of spins $\mathbf{\hat{s}}(\mathbf{r}_i)$ (with $\mathbf{r}_i$ the position of the atomic lattice node), we may approximate the derivatives in a given direction by the difference between $\mathbf{\hat{s}}$ at the nearest neighbor on each side of the spin. Let $\mathbf{\hat{u}}_1=\mathbf{\hat{x}}$ and $\mathbf{\hat{u}}_2=\frac{1}{2}\mathbf{\hat{x}}+\frac{\sqrt{3}}{2}\mathbf{\hat{y}}$  the vector of the hexagonal lattice in the $(\mathbf{\hat{x}},\mathbf{\hat{y}})$ plane, $S$ is then given by
\begin{widetext}
\begin{equation}
\begin{split}
S=\frac{1}{4\pi}\sum_i\mathbf{\hat{s}}(\mathbf{r}_i)
\cdot \left[\left(\frac{
 \mathbf{\hat{s}}(\mathbf{r}_i+\mathbf{\hat{u}}_1-\mathbf{\hat{u}}_2)
+\mathbf{\hat{s}}(\mathbf{r}_i+\mathbf{\hat{u}}_2)
-\mathbf{\hat{s}}(\mathbf{r}_i-\mathbf{\hat{u}}_1+\mathbf{\hat{u}}_2)
-\mathbf{\hat{s}}(\mathbf{r}_i-\mathbf{\hat{u}}_2)
}{2}\right)\right.\\
\times
\left.\left(\frac{
 \mathbf{\hat{s}}(\mathbf{r}_i-\mathbf{\hat{u}}_1+\mathbf{\hat{u}}_2)
+\mathbf{\hat{s}}(\mathbf{r}_i+\mathbf{\hat{u}}_2)
-\mathbf{\hat{s}}(\mathbf{r}_i+\mathbf{\hat{u}}_1-\mathbf{\hat{u}}_2)
-\mathbf{\hat{s}}(\mathbf{r}_i-\mathbf{\hat{u}}_2)}{2\sqrt{3}}
\right) \right]
\end{split}
\end{equation}
\end{widetext}
Note that other finite-differences versions of the derivatives can be built (see also an approximate expression in Ref.~\onlinecite{zhang2016}).

Another definition for a discrete lattice of spins is based on the geometrical interpretation of $S$, namely the
fraction of the order parameter unit sphere that is covered by the two-dimensional magnetization texture
\cite{Berg1981,thiaville2003}.
For each triangle of nearest neighbor sites, the associated 3 spins draw a spherical triangle of the unit sphere.
The generalization of $S$ considers that the area of this spherical triangle is covered on the unit sphere.
Unless the three spins project on three points that lie on the same great circle (think of the equator for example),
there is no ambiguity to define the spherical triangle and whether its surface has to be counted positive or negative.
As a result, this definition of discrete $S$ is an integer number for textures that are uniform at infinity (so that
the physical plane is topologically equivalent to the sphere, by the identification of the points at infinity, which allows
using the results from the topological theory of defects)~\cite{Toulouse76,Mermin79,Kleman83}.
Thus, for a collapsing skyrmion structure, this definition of $S$ will jump from 1 to 0 between two successive configurations,
at the moment where the central triangle crosses the equator in the high symmetry case that is studied here.

Note that the simulations are performed on a finite sample with open boundary conditions, so that a small tilt of magnetization occurs at the edges~\cite{rohart2013}.
If the topological number is calculated on the whole simulation zone, non-integer values are then found (see e.g. Ref.~\onlinecite{sampaio2013}), which
is to be expected as the sample is not equivalent to a sphere.
In order to focus on the skyrmion topology, we restrict the calculation on a 36~nm diameter zone (9064 spins) centered on the skyrmion.

\end{document}